\documentclass[%
 reprint,
 superscriptaddress,
 amsmath,amssymb,
 aps,
 prd,
 floatfix
]{revtex4-2}

\usepackage{graphicx}
\usepackage{dcolumn}
\usepackage{bm}
\usepackage[colorlinks=true,allcolors=blue]{hyperref}
\usepackage{cancel}
\usepackage{siunitx}
\usepackage{fancyhdr}
\makeatletter
\renewcommand{\@date}{Received May, 2026; accepted --; published --}
\makeatother

\begin{document}

\preprint{APS/PRD-Template}

\title{Emergent gravity from nonlinear perturbation of spherical accretion with variable adiabatic index}

\author{Rohit Ghosh}
\email{3333372301@visva-bharati.ac.in}
\affiliation{Department of Physics, Visva-Bharati, Santiniketan - 731235, West Bengal, India}
\author{Souvik Ghose}
\email{dr.souvikghose@gmail.com}
\affiliation{HECRC, University of North Bengal, Raja Rammhunpur, West Bengal, 734013, India}
\affiliation{Harish-Chandra Research Institute, HBNI, Chhatnag Road, Jhunsi, Allahabad - 211109, Uttar Pradesh, India}

\author{Biplab Raychaudhuri}
\email{biplabphy@visva-bharati.ac.in}
\affiliation{Department of Physics, Visva-Bharati, Santiniketan - 731235, West Bengal, India}

\author{Apashanka Das}
\email{apashankadas@hri.res.in}
\affiliation{Harish-Chandra Research Institute, HBNI, Chhatnag Road, Jhunsi, Allahabad - 211109, Uttar Pradesh, India}
\affiliation{Homi Bhabha National Institute (HBNI), Training School Complex, Anushakti Nagar, Mumbai,Maharashtra 400094, India}

\author{Tapas K. Das}
\email{tapas@hri.res.in}
\affiliation{Harish-Chandra Research Institute, HBNI, Chhatnag Road, Jhunsi, Allahabad - 211109, Uttar Pradesh, India}
\affiliation{Homi Bhabha National Institute (HBNI), Training School Complex, Anushakti Nagar, Mumbai,Maharashtra 400094, India}

\begin{abstract}
The main aim of the present work is to demonstrate that the analogue gravity phenomena is not an artifact of linear perturbation, rather gravity-like effects emerge through the non linear higher order perturbation of transonic fluid as well. To establish that fact, a spherically accreting astrophysical system has been considered where the hydrodynamic accretion with a relativistic, multi-component equation of state with position dependent adiabatic index onto compact astrophysical objects has been considered. 
rate. By extending the acoustic metric formalism beyond the linear regime, it has been shown that the aforementioned perturbations satisfy a covariant wave equation in an effective acoustic spacetime with non-linear corrections, making the analogue geometry
dynamical. As a consequence, the acoustic horizon can shift (inward or outward), depending on the relative amplitudes of density, temperature, and mass accretion-rate fluctuations. This provides a more realistic framework to investigate the dynamics of the non-linear analogue spacetime  in astrophysically relevant accretion flows.
\end{abstract}

\maketitle
\thispagestyle{plain}

\section{Introduction}
\label{secI}
Emergent/Analogue gravity provides a framework in which perturbations of transonic fluid flows can be described as scalar fields propagating in an effective curved spacetime containing acoustic horizons. Since the pioneering work of Unruh \cite{Unruh1981}, it has been understood that sound waves in moving fluids can mimic several kinematical properties of black hole spacetimes. This idea was further developed by Visser \cite{Visser1998} and later reviewed extensively by Barceló, Liberati and Visser \cite{Barcelo2011}. Such systems allow certain aspects of horizon physics to be studied in fluid dynamical settings. Astrophysical accretion flows provide particularly important analogue systems because they are naturally transonic. The stability of spherical accretion flows under perturbations was first studied by Moncrief \cite{Moncrief1980}, and subsequent works by Bilić \cite{Bilic1999} and Das and collaborators \cite{Das2004,Dasgupta2005,Abraham2006,DasBilic2007} established accreting black hole systems as natural analogue gravity laboratories containing both gravitational and acoustic horizons. Most of these studies, however, considered linear perturbations and simplified thermodynamic descriptions such as polytropic or isothermal equations of state.

Realistic accretion flows near compact objects may involve relativistic temperatures and multi-species composition, requiring more sophisticated thermodynamic descriptions. In this context, relativistic equations of state developed by Ryu and collaborators \cite{Ryu2006} and further applied to accretion problems by Chattopadhyay and collaborators \cite{Chattopadhyay2013,Chattopadhyay2011} provide a more realistic description of relativistic fluid thermodynamics. More recently, Paul \textit{et al.} \cite{Paul2025} investigated analogue gravity phenomena in multicomponent relativistic accretion flows using such equations of state. The adiabatic index $\Gamma$ is essentially a function of the absolute temperature for this case. Another important extension concerns the role of non-linear perturbations. While linear perturbations lead to stationary acoustic geometries, recent work by Fernandes, Maity and Das \cite{PhysRevD.106.025020} has demonstrated that non-linear perturbations can generate dynamical analogue spacetimes in which the effective acoustic metric evolves and the acoustic horizon may shift depending on the relative behaviour of density and mass accretion rate perturbations.

Motivated by these developments, in this work we study non-linear perturbations of the mass accretion rate in spherically symmetric accretion flows described by a relativistic multi-component equation of state. By extending the acoustic metric construction beyond the linear regime, we show that the perturbations satisfy a covariant wave equation in an effective acoustic spacetime with non-linear corrections, thereby rendering the analogue geometry dynamical. We further examine how non-linear effects influence the behaviour of the acoustic horizon, thus extending previous analogue gravity studies toward more realistic relativistic accretion scenarios.

This paper is organized as follows. In Section~\eqref{secII}, we describe the relativistic equation of state and define the associated adiabatic sound speed. In Section~\eqref{secIII}, we derive the acoustic metric and the wave equation arising from the fluid flow. Section~\eqref{secIV} describes the non-linear perturbations in details. Section~\eqref{secV} describes the variation of the acoustic horizon. In Section~\eqref{secVI}, we analyze the stability of standing and travelling waves. In Section~\eqref{secVII}, we estimate the asymptotic sound speed from the wave equation. Section~\eqref{secVIII} includes the graphical analysis and verification of the results with observational data. Finally, Section~\eqref{secIX} summarizes our conclusions.

\section{A brief on multi-species equation of state}
\label{secII}
We consider a multispecies accretion flow consisting of electrons ($e^-$), positrons ($e^+$), and protons ($p^+$) of proportion parametrized by $\xi$. The number density (n) of the accreting matter is given by:
\begin{equation}
\label{eqn1}
n = \sum_i n_i = n_{e^-} + n_{e^+} + n_{p^+},
\end{equation}
where, $n_{e^-}$, $n_{e^+}$, and $n_{p^+}$ are the electron, positron, and
proton number densities, respectively. If we further demand that the
overall charge neutrality is always maintained, we have:
\begin{equation}
\label{eqn2}
n_{e^-} = n_{e^+} + n_{p^+},
\end{equation}
which gives:
\begin{equation}
\label{eqn3}
n = 2n_{e^-}, \qquad n_{e^+} = n_{e^-}(1 - \xi),
\end{equation}
where, $\xi \equiv \dfrac{n_{p^+}}{n_{e^-}}$.\\\\
The mass density is expressed as:
\begin{equation}
\label{eqn4}
\rho = \sum_i n_i m_i = n_{e^-}m_{e^-}\alpha = \rho_{e^-}\alpha,
\end{equation}
where, $\eta = \frac{m_{e^-}}{m_{p^+}}$, $m_{e^-}$, and $m_{p^+}$ are the electron and proton masses, respectively, and $\alpha=2-\xi\left(1-\frac{1}{\eta}\right)$. The EoS for multispecies flow, as given in \cite{2009ApJ...694..492C}, is:
\begin{equation}
\label{eqn5}
\bar{e} = \sum_i e_i = \sum_i n_i m_i c^2 + p_i
\left( \dfrac{9 p_i + 3 n_i m_i c^2}{3 p_i + 2 n_i m_i c^2} \right),
\end{equation}
where, $\bar{e}$ is the energy density.\\\\
The isotropic pressure can be defined with respect to the Boltzmann constant $K_B$ 
and absolute temperature $T$ as:
\begin{equation}
\label{eqn6}
p = \sum_i p_i = 2 n_{e^-} K_B T.
\end{equation}
We define the dimensionless temperature as:
\begin{equation}
\label{eqn7}
\Theta = \frac{K_B T}{m_{e^-} c^2}.
\end{equation}
Using Eq.~\eqref{eqn7}, the expression for the isotropic pressure 
takes the form:
\begin{equation}
p = 2n_{e^-}m_{e^-}c^2\Theta = 2\rho_{e^-}c^2\Theta= 2\frac{\rho}{\alpha}c^2\Theta.
\label{eqn8}
\end{equation}
The EoS given in Eq.~\eqref{eqn5} can be written as:
\begin{equation}
\label{eqn9}
\bar{e} = n_{e^-} m_{e^-}c^2\Xi = \rho_{e^-}c^2\Xi = \frac{\rho}{\alpha}c^2\Xi,
\end{equation}
where, $\Xi = (2 - \xi) \left[ 1 + \Theta \left( \frac{9 \Theta + 3}{3 \Theta + 2} \right) \right] + \xi \left[ \frac{1}{\eta} + \Theta \left(\frac{9 \Theta + \frac{3}{\eta}}{3 \Theta + \frac{2}{\eta}} \right) \right]$.\\\\
The specific enthalpy can be written as:
\begin{equation}
\label{eqn10}
h = \frac{\bar{e} + p}{\rho} = \frac{c^2}{\alpha} (\Xi+ 2\Theta).
\end{equation}
The polytropic index is conveniently given by \cite{2009ApJ...694..492C}:
\begin{equation}
\label{eqn11}
N = \frac{1}{2} \frac{\text{d}\Xi}{\text{d}\Theta}.
\end{equation}\\\\
Similarly, the adiabatic index can be written as:
\begin{equation}
\label{eqn12}
\Gamma=1+\frac{1}{N}.
\end{equation}\\\\
The adiabatic sound speed is then given by:
\begin{equation}
\label{eqn13}
c_s^2=\frac{2\Gamma\Theta}{\alpha}.
\end{equation}

\section{Wave Equation and Inverse Acoustic Metric}
\label{secIII}
The general form of the continuity equation is given by:
\begin{equation}
\label{eqn14}
\dot{\rho}+\overrightarrow{\nabla}.(\rho\overrightarrow{v})=0.
\end{equation}
The Euler equation is given by:
\begin{equation}
\label{eqn15}
\dot{\overrightarrow{v}}+\left(\overrightarrow{v}.\overrightarrow{\nabla}\right)\overrightarrow{v}+\frac{1}{\rho}\overrightarrow{\nabla}p+\overrightarrow{\nabla}\Phi=0
\end{equation}
In these equations, overdots denote time derivatives and 
$\vec{\nabla}$ is the spatial derivative. The density, pressure, and velocity 
of the fluid are respectively denoted by $\rho$, $p$, and $\vec{v}$, 
while $\Phi=\Phi(r)$ is the potential of an external force.
We will consider spherically symmetric and barotropic flows. 
The barotropic condition $p = p(\rho)$ allows us to define a scalar function 
$H(\rho)$ such that $\vec{\nabla} H = \frac{1}{\rho} \vec{\nabla} p$. 
With the sound speed $c_s$ defined by $c_s^2 = \frac{\partial p}{\partial \rho}$, 
we find:
\begin{equation}
\label{eqn16}
\frac{\partial H}{\partial r}=\frac{1}{\rho}\frac{\partial p}{\partial r}\implies\frac{\partial H}{\partial\rho}\frac{\partial\rho}{\partial r}=\frac{1}{\rho}\frac{\partial p}{\partial\rho}\frac{\partial\rho}{\partial r}\implies\frac{\partial H}{\partial\rho}=\frac{c_{\text{s}}^{2}}{\rho}.
\end{equation}
Spherical symmetry implies a spatial dependence on the radial distance and a non-vanishing radial velocity component, which we denote by $v$. While, the fields also depend on time, we assume a time-independent external potential $\Phi=\Phi(r)$. Hence, Eq.~\eqref{eqn14} and \eqref{eqn15} simplify to the following two-dimensional equations:
\begin{equation}
\label{eqn17}
\dot{\rho}+\frac{1}{r^2}(r^2\rho v)^\prime = 0
\end{equation}
\begin{equation}
\label{eqn18}
 \text{and }\dot{v}+vv^{\prime}+H^{\prime}+\Phi^{\prime}=0,
\end{equation}
where, primes denote derivatives with respect to $r$ in Eqs. \eqref{eqn17} and \eqref{eqn18}.\\\\
We define the mass accretion rate:
\begin{equation}
\label{eqn19}
f=\rho vr^{2}
\end{equation}
and consider the system in terms of $f$ and $\rho$.
\begin{equation}
\label{eqn20}
\therefore\,\,v=\frac{f}{\rho r^{2}}.
\end{equation}
In the case of Eq.~\eqref{eqn17}, we find from direct substitution:
\begin{equation}
\label{eqn21}
\dot{\rho}+\frac{f^{\prime}}{r^{2}}=0\text{.}
\end{equation}
Partially differentiating Eq.~\eqref{eqn20} with respect to $t$ and using Eq.~\eqref{eqn21} gives:
\begin{equation}
\label{eqn22}
\dot{v}=\frac{\dot{f}}{\rho r^{2}}-\frac{f\dot{\rho}}{\rho^{2}r^{2}}=\frac{1}{\rho r^{2}}\left(\dot{f}-\frac{f\dot{\rho}}{\rho}\right)=\frac{1}{\rho r^{2}}\left(\dot{f}+\frac{ff^{\prime}}{\rho r^{2}}\right)\text{.}
\end{equation}
In order to arrive at a wave equation, we partially differentiate Eq.~\eqref{eqn18} with respect to $t$ to get:
\begin{align}
\label{eqn23}
&\partial_{t}\dot{v}+\partial_{t}\left(vv^{\prime}\right)+\partial_{t}H^{\prime}+\cancelto{0}{\partial_{t}\Phi^{\prime}}=0\notag\\
\text{or, }&\partial_{t}\left\{ \frac{1}{\rho r^{2}}\left(\dot{f}+\frac{ff^{\prime}}{\rho r^{2}}\right)\right\} +\partial_{t}\left(\frac{1}{2}\partial_{r}v^{2}\right)+\partial_{r}\dot{H}=0\notag\\
&\hspace{5.25cm}\text{[using Eq.~\eqref{eqn22}]}\notag\\
\text{or, }&\partial_{t}\left(\frac{\dot{f}}{\rho r^{2}}\right)+\partial_{t}\left(\frac{ff^{\prime}}{\rho^{2}r^{4}}\right)+\partial_{r}\left(\frac{\partial_t v^2}{2}\right)+\partial_{r}\left(\frac{\partial H}{\partial\rho}\dot{\rho}\right)\notag\\
&\hspace{7cm}=0\notag\\
\text{or, }&\partial_{t}\left(\frac{\dot{f}}{\rho r^{2}}\right)+\partial_{t}\left(\frac{ff^{\prime}}{\rho^{2}r^{4}}\right)+\partial_{r}\left(v\dot{v}\right)+\partial_{r}\left(\frac{c_{\text{s}}^{2}\dot{\rho}}{\rho}\right)=0\notag\\
&\hspace{5.25cm}\text{[using Eq.~\eqref{eqn16}]}\notag\\
\text{or, }&\partial_{t}\left(\frac{\dot{f}}{\rho r^{2}}\right)+\partial_{t}\left(\frac{ff^{\prime}}{\rho^{2}r^{4}}\right)+\partial_{r}\left(\frac{f\dot{f}}{\rho^{2}r^{4}}+\frac{f^{2}f^{\prime}}{\rho^{3}r^{6}}\right)\notag\\
&\hspace{4.75cm}+\partial_{r}\left(-\frac{c_{\text{s}}^{2}f^{\prime}}{\rho r^{2}}\right)=0\notag\\
&\hspace{2.8cm}\text{[using Eqs. \eqref{eqn20}, \eqref{eqn21} and \eqref{eqn22}]}\notag\\
\text{or, }&\partial_{t}\left(\frac{1}{\rho r^{2}}\partial_{t}f\right)+\partial_{t}\left(\frac{f}{\rho^{2}r^{4}}\partial_{r}f\right)+\partial_{r}\left(\frac{f}{\rho^{2}r^{4}}\partial_{t}f\right)\notag\\
&\hspace{3cm}+\partial_{r}\left\{\left(\frac{f^{2}}{\rho^{3}r^{6}}-\frac{c_{\text{s}}^{2}}{\rho r^{2}}\right)\partial_{r}f\right\} =0
\end{align}
By defining the inverse metric components, 
\begin{equation}
\label{eqn24}
g^{tt} = \frac{1}{\rho r^2}, \quad g^{tr} = \frac{f}{\rho^2 r^4} = g^{rt}, \quad
g^{rr} = \frac{f^2}{\rho^3 r^6} - \frac{c_s^2}{\rho r^2},
\end{equation}
we find that Eq.~\eqref{eqn23} takes the suggestive form of the following wave equation:
\begin{equation}
\label{eqn25}
\partial_\mu (g^{\mu \nu} \partial_\nu f) = 0.
\end{equation}
Equation \eqref{eqn25} will describe fluctuations on a curved background only after perturbatively expanding about a stationary solution. The solutions of Eqs. \eqref{eqn17} and \eqref{eqn18} are generally not the same as those for Eqs. \eqref{eqn21} and \eqref{eqn25}, since Eq.~\eqref{eqn25} results from a time derivative of Eq.~\eqref{eqn18}. However, boundary conditions specify a unique stationary solution of eqs. \eqref{eqn17} and \eqref{eqn18} that manifestly satisfy Eqs. \eqref{eqn21} and \eqref{eqn25}. Subsequently, a perturbative solution for $f$ from the original fluid equations will be one of the solutions of Eq.~\eqref{eqn25} through a consistently chosen boundary condition in time.\\\\
We can further express Eq.~\eqref{eqn25} as [see \cite{Visser1998}]:  
\begin{equation}
\label{eqn26}
\frac{1}{\sqrt{-G}}\partial_\mu\left(\sqrt{-G}G^{\mu\nu}\partial_\nu f\right) = 0,
\end{equation}
where, $\sqrt{-G}G^{\mu\nu}=g^{\mu\nu}$.
For two-dimensional flows, this is achieved by introducing additional spatial metric components, such as $g^{\theta\theta}$ and $g^{\phi\phi} = \frac{g^{\theta\theta}}{\sin^2\theta}$ that respect spherical symmetry without modifying Eq.~\eqref{eqn25}.  As this transformation does not affect the causal structure or dynamics, we will consider the unique two-dimensional inverse effective metric $g^{\mu\nu}$ with components given in Eq.~\eqref{eqn24}.

\section{Non-linear perturbations}
\label{secIV}
The flow governed by Eqs. \eqref{eqn21} and \eqref{eqn25} can be 
perturbatively solved about a spherically symmetric stationary solution, 
characterized by a constant mass accretion rate $f_0$ and a time-independent 
density $\rho_0(r)$. In this case, an $n$th order expansion of 
$f(r,t)$ and $\rho(r,t)$ takes the form:
\begin{equation}
\label{eqn27}
f(r,t) = f_0 + \sum_{i=1}^{n} \epsilon^i f_i(r,t),
\end{equation}
\begin{equation}
\label{eqn28}
\rho(r,t) = \rho_0(r) + \sum_{i=1}^{n} \epsilon^i \rho_i(r,t),
\end{equation}
where, $\epsilon$ is a dimensionless counting parameter, whose power 
identifies the perturbation order. If a general field $A(r,t)$ constructed 
from $\rho$ and $f$ has the expansion  
\begin{equation}
\label{eqn29}
A(r,t) = A_0(r) + \sum_{i=1}^{n} \epsilon^i A_i(r,t),
\end{equation}
then it is perturbative provided  
\begin{equation}
\label{eqn30}
\epsilon \frac{|A_{l+1}|}{|A_l|} < 1, \qquad (l = 0,\,\ldots,\,n-1).
\end{equation}
Substituting Eqs. \eqref{eqn27} and \eqref{eqn28} in Eq.~\eqref{eqn21} gives:
\begin{align}
\label{eqn31}
&\sum_{i=1}^{n} \epsilon^i \dot{\rho}_i(r,t) + \frac{1}{r^2} \sum_{i=1}^{n} \epsilon^i f_i'(r,t) = 0\notag\\
\text{or, } &\sum_{i=1}^{n} \epsilon^i \left\{\dot{\rho}_i(r,t) + \frac{1}{r^2} f_i'(r,t) \right\} = 0\notag\\
\text{or, } &\dot{\rho}_i(r,t) + \frac{1}{r^2} f_i'(r,t) = 0
\end{align}
It follows from Eq.~\eqref{eqn31} that:
\begin{equation}
\label{eqn32}
\dot{\rho}_1 + \frac{\partial_r f_1}{r^2} = 0
\end{equation}
\begin{equation}
\label{eqn33}
\text{and }\dot{\rho}_2 + \frac{\partial_r f_2}{r^2} = 0.
\end{equation}
Substituting Eq.~\eqref{eqn27} in Eq.~\eqref{eqn25} and perturbing $g^{\mu\nu}$ non-linearly gives:
\begin{align}
\label{eqn34}
&\partial_\mu \left[\left\{ g^{\mu\nu}_{(0)} + \sum_{i=1}^{n} \epsilon^i g^{\mu \nu}_{(i)}\right\} \partial_\nu \left( f_0 + \sum_{i=1}^{n} \epsilon^i f_i \right)\right] = 0\notag\\
\text{or, }&\partial_\mu\left[\left\{g^{\mu\nu}_{(0)}+\sum_{i=1}^{n}\epsilon^i g^{\mu \nu}_{(i)}\right\}\left(\sum_{i=1}^{n}\epsilon^i\partial_\nu f_i\right)\right] = 0,
\end{align}\\
where,
\begin{equation}
\label{eqn35}
g_{(0)}^{tt}=\frac{1}{\rho_{0}r^{2}},\, g_{(0)}^{tr}=g_{(0)}^{rt}=\frac{f_{0}}{\rho_{0}^{2}r^{4}}\,\text{ and }\, g_{(0)}^{rr}=\frac{f_{0}^{2}}{\rho_{0}^{3}r^{6}}-\frac{c_{s0}^{2}}{\rho_{0}r^{2}}.
\end{equation}
Here, $c_{s0}^2 = \dfrac{2\Gamma_0\Theta_0}{\alpha}.$\\\\
The coefficient of $\epsilon$ in Eq.~\eqref{eqn34} is:
\begin{equation}
\label{eqn36}
\partial_{\mu}\left\{ g_{(0)}^{\mu\nu}\partial_{\nu}f_{1}\right\}=0.
\end{equation}
The 1st order fluctuation $f_1(r,t)$ propagates on an acoustic background constructed entirely from the background flow. This agrees with known approaches to stationary analogue spacetimes. We can solve Eqs. \eqref{eqn32} and \eqref{eqn36} for $f_1(r,t)$ and $\rho_1(r,t)$.\\\\
The coefficient of $\epsilon^2$ in Eq.~\eqref{eqn34} is:
\begin{equation}
\label{eqn37}
\partial_{\mu}\left\{g_{(0)}^{\mu\nu}\partial_{\nu}f_{2}\right\} +\partial_{\mu}\left\{g_{(1)}^{\mu\nu}\partial_{\nu}f_{1}\right\}=0.
\end{equation}
In order to find out the $g_{(1)}^{\mu\nu}$ components, we substitute Eqs. \eqref{eqn27} and \eqref{eqn28} in Eqs. \eqref{eqn24} to get:
\begin{align}
\label{eqn38}
g^{tt}&= \frac{1}{\left\{ \rho_0(r) + \displaystyle\sum_{i=1}^{n} \epsilon^i \rho_i(r,t) \right\} r^2}\notag\\
&= \frac{1}{\rho_0(r) r^2}
\left\{ 1 + \sum_{i=1}^{n} \epsilon^i \frac{\rho_i(r,t)}{\rho_0(r)} \right\}^{-1}\notag\\
&= \frac{1}{\rho_0(r) r^2} \left\{1 - \sum_{i=1}^{n} \epsilon^i \frac{\rho_i(r,t)}{\rho_0(r)} + \cdots \right\}\notag\\
&= g_{(0)}^{tt} - \sum_{i=1}^{n} \epsilon^i \frac{\rho_i(r,t)}{\{\rho_0(r)\}^2 r^2} + \cdots,
\end{align}
\begin{align}
\label{eqn39}
g^{tr}&=g^{rt}\notag\\
&=\frac{f_0 + \displaystyle\sum_{i=1}^{n} \epsilon^i f_i(r,t)}{\left\{\rho_0(r) + \displaystyle\sum_{i=1}^{n} \epsilon^i \rho_i(r,t) \right\}^2 r^4}\notag\\
&=\left[\frac{f_0 + \displaystyle\sum_{i=1}^{n} \epsilon^i f_i(r,t)}{\{\rho_0(r)\}^2 r^4}
\right] \left\{1 - 2 \displaystyle\sum_{i=1}^{n} \epsilon^i \frac{\rho_i(r,t)}{\rho_0(r)} + \cdots\right\}\notag\\
&=g_{(0)}^{tr} + \displaystyle\sum_{i=1}^{n} \epsilon^i \left[\frac{f_i(r,t)}{\left\{\rho_0(r)\right\}^2 r^4}- \frac{2 f_0 \rho_i(r,t)}{\left\{\rho_0(r)\right\}^3 r^4}\right]+ \cdots
\end{align}
and
\begin{align}
\label{eqn40}
&\qquad g^{rr}\notag\\
&=\frac{\left\{f_0+\displaystyle\sum_{i=1}^n\epsilon^i f_i(r,t)\right\}^2}{\left\{\rho_0(r) + \displaystyle\sum_{i=1}^{n} \epsilon^i \rho_i(r,t) \right\}^3 r^6}\notag\\
&-\frac{c_s^2(\Theta_0) + \left(\displaystyle\sum_{i=1}^{n}\epsilon^i\Theta_i\right)\left. \dfrac{\partial c_s^2}{\partial \Theta} \right|_{\Theta_0}+\dfrac{\left(\displaystyle\sum_{i=1}^{n} \epsilon^i \Theta_i \right)^2 \left. \dfrac{\partial^2 c_s^2}{\partial \Theta^2} \right|_{\Theta_0}}{2!}+\cdots}{\left\{\rho_0(r) + \displaystyle\sum_{i=1}^{n} \epsilon^i \rho_i(r,t) \right\} r^2}\notag\\
&=\left(\frac{f_0^2 + 2 \displaystyle\sum_{i=1}^{n} \epsilon^i f_0 f_i + \cdots}{\rho_0^3 r^6}\right)\left\{1-3\displaystyle\sum_{i=1}^{n} \epsilon^i \frac{\rho_i}{\rho_0}+\cdots\right\}\notag\\
&-\left\{\frac{c_{s_0}^2+ \left( \displaystyle\sum_{i=1}^{n} \epsilon^i \Theta_i \right)
\left.\dfrac{\partial c_s^2}{\partial \Theta} \right|_{\Theta_0}+ \cdots}{\rho_0 r^2}
\right\}\left(1-\displaystyle\sum_{i=1}^{n} \epsilon^i \frac{\rho_i}{\rho_0}+\cdots\right)\notag\\
&=g_{(0)}^{rr}+\sum_{i=1}^{n} \epsilon^i\left(\frac{2 f_0 f_i}{\rho_0^3 r^6}- \frac{3 f_0^2 \rho_i}{\rho_0^4 r^6}- \frac{\Theta_i\left.\frac{\partial c_s^2}{\partial\Theta} \right|_{\Theta_0}}{\rho_0 r^2}+ \frac{c_{s_0}^2 \rho_i}{\rho_0^2 r^2}\right)+ \cdots
\end{align}
From the coefficients of $\epsilon$ in Eqs. \eqref{eqn38}-\eqref{eqn40} we get:
\begin{equation}
\label{eqn41}
g_{(1)}^{tt}=-\frac{\rho_1}{\rho_0^2 r^2}=\frac{1}{r^{2}\rho_{0}}\left(-\frac{\rho_{1}}{\rho_{0}}\right),
\end{equation}
\begin{equation}
\label{eqn42}
g_{(1)}^{tr}=g_{(1)}^{rt}=\frac{f_1}{\rho_0^2r^4}-\frac{2f_0\rho_1}{\rho_0^3 r^4}=\frac{f_{0}}{r^{4}\rho_{0}^{2}}\left(\frac{f_{1}}{f_{0}}-2\frac{\rho_{1}}{\rho_{0}}\right)
\end{equation}
\begin{align}
\label{eqn43}
\text{and }g_{(1)}^{rr}&=\frac{2 f_0 f_1}{\rho_0^3 r^6}- \frac{3 f_0^2 \rho_1}{\rho_0^4 r^6}
- \frac{\Theta_1 \left.\frac{\partial c_s^2}{\partial\Theta}\right|_{\Theta_0}}{\rho_0 r^2} +\frac{c_{s_0}^2 \rho_1}{\rho_0^2 r^2}\notag\\
&=\frac{f_0^2}{r^6\rho_0^3}\left(2\frac{f_1}{f_0}-3\frac{\rho_1}{\rho_0}\right)-\frac{c_{s_0}^2}{\rho_0 r^2}\left(\frac{\Theta_1}{c_{s_0}^2}\left.\frac{\partial c_s^2}{\partial\Theta}\right|_{\Theta_0}-\frac{\rho_1}{\rho_0}\right),
\end{align}
respectively.\\\\
Partially differentiating Eq.~\eqref{eqn13} with respect to $\Theta$ gives:
\begin{align*}
&\frac{\partial c_s^2}{\partial\Theta} = \frac{2\Gamma}{\alpha}+\frac{2\Theta}{\alpha}\frac{\text{d}\Gamma}{\text{d}\Theta}=\frac{1}{\alpha}\left(2\Gamma-\frac{\Theta}{N^2}\frac{\text{d}^2\Xi}{\text{d}\Theta^2}\right)\\
\text{or, }&\frac{1}{c_{s_0}^2}
\left.\frac{\partial c_s^2}{\partial\Theta}\right|_{\Theta_0}=\frac{1}{\Theta_0}-\frac{(\Gamma_0-1)^2}{2\Gamma_0}\left.\frac{\text{d}^2\Xi}{\text{d}\Theta^2}\right|_{\Theta_0}\\
\text{or, }&\frac{1}{c_{s_0}^2}
\left.\frac{\partial c_s^2}{\partial\Theta}\right|_{\Theta_0}=\frac{1}{\Theta_0}+\frac{1}{\Gamma_0}\left.\frac{\text{d}\Gamma}{\text{d}\Theta}\right|_{\Theta_0}.
\end{align*}
Hence, Eq.~\eqref{eqn43} boils down to:
\begin{align}
\label{eqn44}
&g_{(1)}^{rr}\notag\\
=&\frac{f_0^2}{r^6\rho_0^3}\left(2\frac{f_1}{f_0}-3\frac{\rho_1}{\rho_0}\right)-\frac{c_{s_0}^2}{\rho_0 r^2}\left(\frac{\Theta_1}{\Theta_0}+\frac{1}{\Gamma_0}\left.\frac{\text{d}\Gamma}{\text{d}\Theta}\right|_{\Theta_0}-\frac{\rho_1}{\rho_0}\right)\\
=&\frac{f_0^2}{r^6\rho_0^3}\left(2\frac{f_1}{f_0}-3\frac{\rho_1}{\rho_0}\right)-\frac{2\Gamma_0\Theta_0}{\alpha\rho_0 r^2}\left(\frac{\Theta_1}{\Theta_0}+\frac{1}{\Gamma_0}\left.\frac{\text{d}\Gamma}{\text{d}\Theta}\right|_{\Theta_0}-\frac{\rho_1}{\rho_0}\right).
\end{align}\\
Eq.~\eqref{eqn25} informs us that Eqs. \eqref{eqn36} and \eqref{eqn37} are collectively a second-order fluctuation $f_0 + \epsilon f_1 + \epsilon^2 f_2$ propagating on an effective first-order background with inverse metric $g^{\mu \nu}_{(0)} + \epsilon g^{\mu \nu}_{(1)}$.\\
The iterative process can be carried out to all orders. At order $n$, we have from Eq.~\eqref{eqn31}:
\begin{equation}
\label{eqn46}
\dot{\rho}_n=-\frac{\partial_r f_n}{r^2}.
\end{equation}
Eq.~\eqref{eqn37} can be rewritten as:
\begin{equation}
\label{eqn47}
\partial_{\mu}\left\{g_{(0)}^{\mu\nu}\partial_{\nu}f_{2}\right\} =-\partial_{\mu}\left\{g_{(1)}^{\mu\nu}\partial_{\nu}f_{1}\right\}.
\end{equation}
Equating the coefficient of $\epsilon^3$ in Eq.~\eqref{eqn34} gives:
\begin{align}
\label{eqn48}
&\partial_\mu\left\{g^{\mu\nu}_{(0)}\partial_\nu f_3\right\}+\partial_\mu\left\{g^{\mu\nu}_{(1)} \partial_\nu f_2 \right\}+\partial_\mu\left\{g^{\mu\nu}_{(2)}\partial_\nu f_1\right\}=0\notag\\
\text{or, }&\partial_\mu\left\{g^{\mu\nu}_{(0)}\partial_\nu f_3\right\}=-\partial_\mu\left\{g^{\mu\nu}_{(2)}\partial_\nu f_1+g^{\mu\nu}_{(1)} \partial_\nu f_2 \right\}.
\end{align}
Hence, at order $n$ we have:
\begin{equation}
\label{eqn49}
\partial_\mu\left\{g^{\mu\nu}_{(0)}\partial_\nu f_n\right\}=-\sum_{i=1}^{n-1}\partial_\mu\left\{g^{\mu\nu}_{(i)}\partial_\nu f_{n-i}\right\}.
\end{equation}
The $n$th order fluctuation $f=f_0+\displaystyle\sum_{i=1}^{n}\epsilon^i f_i=f_0+\cdots+\epsilon^n f_n$ propagates on an effective $(n-1)$th order background with inverse metric:
\begin{align}
\label{eqn50}
&g^{\mu\nu}_{(0)}+\epsilon g^{\mu\nu}_{(1)}+\epsilon^2 g^{\mu\nu}_{(2)}+\cdots+\epsilon^{n-1}g^{\mu\nu}_{(n-1)}=\sum_{i=0}^{n-1}\epsilon^i g^{\mu\nu}_{(i)}\notag\\
\text{or, }&g^{\mu\nu}_{\text{eff}(n-1)}=\sum_{i=0}^{n-1}\epsilon^i g^{\mu\nu}_{(i)}.
\end{align}

\section{Acoustic Metric and Horizon Variations}
\label{secV}
The inverse metric to all orders can be compactly represented using Eq.~\eqref{eqn24}, with the order $n$ expression resulting from Eq.~\eqref{eqn50}. Inverting Eq.~\eqref{eqn24} provides the following acoustic metric:
\begin{align}
\label{eqn51}
[g_{\mu\nu}]
=
\begin{bmatrix}
g_{tt} & g_{tr} \\
g_{rt} & g_{rr}
\end{bmatrix}
&=
\begin{bmatrix}
gg^{rr} & -gg^{rt} \\
-gg^{tr} & gg^{tt}
\end{bmatrix}\notag\\
&=
\begin{bmatrix}
\rho r^{2}(1-\beta^{2}) & \dfrac{f}{c_s^{2}} \\
\dfrac{f}{c_s^{2}} & -\dfrac{\rho r^{2}}{c_s^{2}}
\end{bmatrix},
\end{align}
where, $g = -\dfrac{\rho^2 r^2}{c_s^2}= -\dfrac{\alpha\rho^2 r^2}{2\Gamma\Theta}$ is the determinant of the metric and
$\beta = \dfrac{f}{c_s \rho r^{2}}$ is the fluid velocity to sound speed ratio.
In units with the speed of light $c=1$, we have $0 < v = \dfrac{f}{\rho r^{2}} < 1$ and $0 < \beta < \dfrac{1}{c_s}$. The condition $g_{tt} = 0$, or equivalently $g^{rr} = 0$, determines the horizon radius $r_H$
\begin{equation}
\label{eqn52}
r_H^4=\frac{f^2}{\rho^2 c_s^2}=\frac{\alpha f^2}{2\Gamma\Theta\rho^2}.
\end{equation}
Taking logarithmic derivative of Eq.~\eqref{eqn52} gives:
\begin{align}
\label{eqn53}
&4\frac{\delta r_H}{r_H}=2\frac{\delta f}{f}-2\frac{\delta\rho}{\rho}-\frac{\delta c_s^2}{c_s^2}\notag\\
\text{or, }&\frac{\delta r_H}{r_H}=\frac{1}{2}\frac{\delta f}{f}-\frac{1}{2}\frac{\delta\rho}{\rho}-\frac{1}{4}\frac{\delta \Theta}{c_s^2}\frac{\partial c_s^2}{\partial \Theta}\notag\\
&\hspace{0.67cm}=\frac{1}{2}\frac{\delta f}{f}-\frac{1}{2}\frac{\delta\rho}{\rho}-\frac{1}{4}\left(\frac{1}{\Theta}+\frac{1}{\Gamma}\frac{\text{d}\Gamma}{\text{d}\Theta}\right)\delta\Theta\notag\\
&\hspace{0.67cm}=\frac{1}{2}\left[\frac{\delta f}{f}-\left[\frac{\delta\rho}{\rho}+\frac{1}{2}\left(\frac{1}{\Theta}+\frac{1}{\Gamma}\frac{\text{d}\Gamma}{\text{d}\Theta}\right)\delta\Theta\right]\right].
\end{align}
Here, $f$, $\rho$, $\Theta$ and $r_H$ denote values at a specific order in perturbation, while $\delta f$, $\delta\rho$, $\delta \Theta$ and $\delta r_H$ describe their relative variations to the next order. Using Eq.~\eqref{eqn21}, we deduce:
\begin{align}
\label{eqn54}
&\dot{\rho}+\frac{f^{\prime}}{r^{2}}=0\notag\\
\text{or, }&f^{\prime}=-\dot{\rho}r^2\notag\\
\text{or, }&f=-\int\text{d}r r^2\dot{\rho}\notag\\
\text{or, }&\frac{\delta f}{f}=\frac{\int\text{d}r r^2\delta\dot{\rho}}{\int\text{d}r r^2\dot{\rho}}
\end{align}
The relative mass accretion rate change $\frac{\delta f}{f}$ is thus related to a spatially averaged change in energy and is expected to be positive. The relative density change $\frac{\delta\rho}{\rho}$ and the relative temperature change $\frac{\delta\Theta}{\Theta}$ however are not constrained. In particular, Eq.~\eqref{eqn53} leads to the following two cases:
\begin{equation}
\label{eqn55}
\frac{\delta\rho}{\rho}+\frac{1}{2}\left(\frac{1}{\Theta}+\frac{1}{\Gamma}\frac{\text{d}\Gamma}{\text{d}\Theta}\right)\delta\Theta>\frac{\delta f}{f}
\end{equation}
for a receding acoustic horizon and
\begin{equation}
\label{eqn56}
\frac{\delta\rho}{\rho}+\frac{1}{2}\left(\frac{1}{\Theta}+\frac{1}{\Gamma}\frac{\text{d}\Gamma}{\text{d}\Theta}\right)\delta\Theta<\frac{\delta f}{f}
\end{equation}
for an advancing acoustic horizon.

\section{Stability Analysis of Standing and Travelling Waves}
\label{secVI}
Since the explicit radial and temporal dependency of the acoustic metric components ($g_{\mu\nu}$) is not known in general, we now focus exclusively on the stability analysis of $1^{\text{st}}$ order perturbations.
Consider a trial wave function as a solution to the $1^{\text{st}}$ order wave equation [Eq.~\eqref{eqn36}]:
\begin{equation}
\label{eqn57}
f_1 = \tilde{f}_{\omega}(r)e^{-i\omega t}.
\end{equation}
Therefore, Eq.~\eqref{eqn36} becomes:
\begin{align}
\label{eqn58}
&\partial_t\left[g^{tt}_{(0)}\,\partial_t f_1\right]+ \partial_t\left[g^{tr}_{(0)}\,\partial_r f_1\right]+ \partial_r\left[g^{rt}_{(0)}\,\partial_t f_1\right]\notag\\
&\hspace{5cm}+ \partial_r\left[g^{rr}_{(0)}\,\partial_r f_1\right] = 0\notag\\
\text{or, }&g^{tt}_{(0)}\,\partial_t^2 f_1+g^{tr}_{(0)}\,\partial_t \partial_r f_1 + \partial_r g^{rt}_{(0)}\,\partial_t f_1 + g^{rt}_{(0)}\,\partial_r \partial_t f_1\notag\\
&\hspace{3.65cm}+ \partial_r g^{rr}_{(0)}\,\partial_r f_1 + g^{rr}_{(0)}\,\partial_r^2 f_1 = 0\notag\\
\text{or, }&\omega^2 g^{tt}_{(0)}\tilde f_\omega + 2 i\omega g^{tr}_{(0)}\tilde f'_\omega + i\omega \left[\partial_r g^{rt}_{(0)}\right]\tilde f_\omega - \left[\partial_r g^{rr}_{(0)}\right]\tilde f'_\omega\notag\\
&\hspace{6.05cm}- g^{rr}_{(0)}\tilde f''_\omega = 0\notag\\
\text{or, }&\omega^2 g^{tt}_{(0)}\tilde f_\omega  + i\omega\left[2 g^{tr}_{(0)}\tilde f'_\omega + \tilde f_\omega\partial_r g^{rt}_{(0)}\right]-[\partial_r g^{rr}_{(0)}]\tilde f'_\omega\notag\\
&\hspace{6.1cm} - g^{rr}\tilde f''_\omega = 0.
\end{align}

\subsection{Standing Wave Analysis}
\label{subsecVI.A}
If the accreting object has a clearly defined physical surface (such as a neutron star), it is expected that any perturbation must vanish both at large distances from the object and at its boundary. In such cases, a standing-wave type solution becomes relevant. However, in a transonic flow, once the fluid crosses the sonic point and becomes supersonic, it can return to a subsonic state only through a discontinuous shock. Since a standing wave represents a continuous solution, this situation is incompatible with transonic behaviour. Therefore, for a standing-wave solution to exist, the flow must remain subsonic everywhere.\\\\
We will now solve the quadratic equation in $\omega$, given by Eq.~\eqref{eqn58}. Let the two boundaries be $r_1$ and $r_2$ ($>r_1$), such that:
\begin{equation}
\label{eqn59}
\tilde{f}_{\omega}(r_1)=\tilde{f}_{\omega}(r_2)=0.
\end{equation}
Multiplying Eq.~\eqref{eqn58} by $\tilde{f}_{\omega}(r)$ and integrating  w.r.t. $r$ from $r_1$ to $r_2$ gives:
\begin{align}
\label{eqn60}
&\omega^2 \int_{r_1}^{r_2} g^{tt}_{(0)}\tilde {f}_\omega^{2}\text{d}r+ i\omega\left[2\int_{r_1}^{r_2}
g^{tr}_{(0)}\tilde f'_\omega \tilde f_\omega \text{d}r+\int_{r_1}^{r_2}\tilde f_\omega^{2}\partial_r g^{rt}_{(0)}\text{d}r\right]\notag\\
&\hspace{1.8cm}- \int_{r_1}^{r_2} \left[\partial_r g^{rr}_{(0)}\right]\tilde f_\omega \tilde f'_\omega\text{d}r -\int_{r_1}^{r_2} g^{rr}_{(0)}\tilde {f}''_\omega \tilde f_\omega\text{d}r = 0\notag\\
\text{or, }&\omega^2 \int_{r_1}^{r_2} g^{tt}_{(0)}\tilde {f}_\omega^{2}\text{d}r-\int_{r_1}^{r_2}\tilde f_\omega\partial_r\left[g^{rr}_{(0)}\tilde {f}'_\omega\right]\text{d}r\notag\\
&\hspace{4.5cm}+i\omega\cancelto{0}{\int_{r_1}^{r_2} \partial_r\left[\tilde f_\omega^{2} g^{rt}_{(0)}\right]\text{d}r}= 0\notag\\
\text{or, }&\omega^2 \int_{r_1}^{r_2} g^{tt}_{(0)}\tilde {f}_\omega^{2}\text{d}r-\cancelto{0}{\Big[\tilde f_\omega g^{rr}_{(0)}\tilde f'_\omega \Big]_{r_1}^{r_2}}+ \int_{r_1}^{r_2} g^{rr}_{(0)}\,(\tilde f'_\omega)^2dr= 0\notag\\
\text{or, }&\omega^2 = -\frac{\displaystyle\int_{r_1}^{r_2}g^{rr}_{(0)}(\tilde f'_\omega)^2\text{d}r}{\displaystyle\int_{r_1}^{r_2}g^{tt}_{(0)}\tilde f_\omega^2\text{d}r}.
\end{align}
Now, $g^{tt}_{(0)}=\dfrac{v_0}{f_0}$ and $g^{rr}_{(0)}=\dfrac{v_0(v_0^2-c_{s0}^2)}{f_0}$. Therefore, Eq.~\eqref{eqn60} becomes:
\begin{equation}
\label{eqn61}
\omega^2 = -\frac{\displaystyle\int_{r_1}^{r_2}v_0(v_0^2-c_{s0}^2)(\tilde f'_\omega)^2\text{d}r}{\displaystyle\int_{r_1}^{r_2}v_0\tilde f_\omega^2\text{d}r}.
\end{equation}
Since the flow is subsonic everywhere ($v_0<c_{s0}$), $\omega$ is necessarily real in this region, leading to an oscillatory standing-wave solution even in the presence of a divergent term. Hence, the standing waves are stable under linear perturbations.
\subsection{Travelling Wave Analysis}
\label{subsecVI.B}
If the accretor is a black hole, there is no solid surface, and any flow reaching the event horizon becomes supersonic. In this case, the analysis presented in the previous subsection no longer applies. The wave behaves as a travelling wave, with a wavelength much smaller than the radius of the accretor. Employing the Wentzel--Kramers--Brillouin (WKB) approximation method, we can approximate the solution of Eq.~\eqref{eqn58} in the following power series of $\omega$:
\begin{equation}
\label{eqn62}
\tilde f_\omega(r) = \exp\left[\sum_{n=-1}^{\infty}\frac{k_n(r)}{\omega^{n}}\right].
\end{equation}
Therefore, $\partial_r\tilde f_\omega=\tilde f_\omega\displaystyle\sum_{n=-1}^{\infty}\frac{k_n'(r)}{\omega^n}$
and
\begin{align*}
\partial_r^2 \tilde f_\omega &= (\partial_r \tilde f_\omega)\sum_{n=-1}^{\infty}\frac{k_n'(r)}{\omega^n}+\tilde f_\omega\sum_{n=-1}^{\infty}\frac{k_n''(r)}{\omega^n}\notag\\
&=\tilde f_\omega\left[\left\{\sum_{n=-1}^{\infty}\frac{k_n'(r)}{\omega^{n}}\right\}
\left\{\sum_{l=-1}^{\infty}\frac{k_l'(r)}{\omega^l}\right\}+\sum_{n=-1}^{\infty}\frac{k_n''(r)}{\omega^{n}}\right].
\end{align*}
Hence, Eq.~\eqref{eqn58} becomes:
\begin{align}
\label{eqn63}
&\omega^2 g^{tt}_{(0)}+i\omega\left\{2g^{tr}_{(0)}\sum_{n=-1}^{\infty} \frac{k_n'(r)}{\omega^{n}}+\partial_r g^{rt}_{(0)}\right\}\notag\\
&-\left[\partial_r g^{rr}_{(0)}\right]
\sum_{n=-1}^{\infty} \frac{k_n'(r)}{\omega^{n}}\notag\\
&-g^{rr}_{(0)}\left[\left\{\sum_{n=-1}^{\infty}\frac{k_n'(r)}{\omega^{n}}\right\}\left\{\sum_{l=-1}^{\infty}\frac{k_l'(r)}{\omega^l}\right\}+\sum_{n=-1}^{\infty}\frac{k_n''(r)}{\omega^{n}}\right]=0.
\end{align}
Equating the coefficients of $\omega^2$, $\omega$, $\omega^0$ and $\omega^{-1}$ in Eq.~\eqref{eqn63} give:
\begin{align}
\label{eqn64}
&g^{tt}_{(0)} + 2 i g^{tr}_{(0)} k'_{-1} - g^{rr}_{(0)}(k'_{-1})^2=0\notag\\
\text{or, }&\dfrac{v_0(v_0^2 - c_{s0}^2)}{f_0}(k'_{-1})^2 - 2i\frac{v_0^2}{f_0}\, k'_{-1} - \frac{v_0}{f_0}=0\notag\\
\text{or, }&\bigl\{(v_0 + c_{s0})k'_{-1} - i\bigr\}\bigl\{(v_0 - c_{s0})k'_{-1} - i\bigr\}=0\notag\\
\text{or, }&k'_{-1}=\frac{i}{v_0 \pm c_{s0}}\notag\\
\text{or, }&k_{-1}=i\int\frac{\text{d}r}{v_0\pm c_{s0}},
\end{align}

\begin{align}
\label{eqn65}
i\left[2 g^{tr}_{(0)} k'_0+\partial_r g^{rt}_{(0)}\right]&-\left[\partial_r g^{rr}_{(0)}\right]k'_{-1}\notag\\
&-g^{rr}_{(0)}\left(2 k'_{-1} k'_0 + k''_{-1}\right)=0\notag\\
\text{or, }\left[2ig^{tr}_{(0)}-2g^{rr}_{(0)}k'_{-1}\right]k'_0=&\left[k'_{-1}\partial_r g^{rr}_{(0)}+k''_{-1}g^{rr}_{(0)}\right]\notag\\
&-i\partial_r g^{rt}_{(0)}\notag\\
\text{or, }k'_0=-\frac{\partial_r\left[k'_{-1}g^{rr}_{(0)}-ig^{tr}_{(0)}\right]}{2\left[k'_{-1}g^{rr}_{(0)}-ig^{tr}_{(0)}\right]}\notag\\
\text{or, }k'_0=-\frac{1}{2}\partial_r\left[\ln\left(\frac{c_{s0}}{\rho_0r^2}\right)\right]\notag\\
\text{or, }k_0=-\frac{1}{2}\ln\left(\frac{c_{s0}}{\rho_0r^2}\right),
\end{align}
\begin{align}
\label{eqn66}
&2ig^{tr}_{(0)}k'_1-\left[\partial_r g^{rr}_{(0)}\right]k'_0-g^{rr}_{(0)}\left[2 k'_{-1} k'_1+(k'_0)^2+k''_0\right]=0\notag\\
\text{or, }&\pm i\frac{8c_{s0}}{\rho_0r^2}k'_1-2\partial_r\left[\left(\frac{v_0^2-c_{s0}^2}{\rho_0r^2}\right)\partial_r\left[\ln\left(\frac{c_{s0}}{\rho_0}\right)-2\ln r\right]\right]\notag\\
&\hspace{1.35cm}+\left[\partial_r\left[\ln\left(\frac{c_{s0}}{\rho_0}\right)-2\ln r\right]\right]^2\left(\frac{v_0^2-c_{s0}^2}{\rho_0r^2}\right)=0\notag\\
\text{or, }&k'_1=\pm \frac{i}{c_{s\infty}r^2}\notag\\
\text{or, }&k_1=\mp \frac{i}{c_{s\infty}r}
\end{align}
and
\begin{align}
\label{eqn67}
&-2ig^{tr}_{(0)}k'_2+k'_1\partial_r g^{rr}_{(0)}+g^{rr}_{(0)}\left(k'_{-1}k'_2+2k'_0k'_1+k''_1\right)=0\notag\\
\text{or, }&\left(\frac{-v_0\pm c_{s0}}{\rho_0r^2}\right)k'_2\pm\partial_r\left(\frac{v_0^2-c_{s0}^2}{\rho_0r^4}\right)\notag\\
&\hspace{1.95cm}\mp\left(\frac{v_0^2-c_{s0}^2}{\rho_0r^4}\right)\partial_r\left[\ln\left(\frac{c_{s0}}{\rho_0}\right)-2\ln r\right]=0\notag\\
\text{or, }&k'_2=\mp \frac{2c_{s\infty}}{r^3}\notag\\
\text{or, }&k_2=\pm \frac{c_{s\infty}}{r^2},
\end{align}
respectively. Here we have evaluated Eqs.~\eqref{eqn66} and \eqref{eqn67} in the asymptotic limit: $\rho_0\big|_{r\to\infty}\rightarrow\rho_\infty$,\,
$c_{s0}\big|_{r\to\infty}\rightarrow c_{s\infty}$ and $v_0\big|_{r\to\infty}\rightarrow 0$.\\\\
Therefore, it follows that:
\begin{equation}
\label{eqn68}
k_{-1}\sim ir,\,k_0\sim\ln r,\,k_1\sim\frac{i}{r}\text{ and } k_2\sim\frac{1}{r^2}.
\end{equation}
Hence, the odd and the even modes are phase related and amplitude related, respectively.\\\\
We see that, for $\omega\gg1$:
\begin{align}
\label{eqn69}
&\omega^2|k_{-1}|\gg\omega|k_0|\gg\omega^0|k_1|\gg\omega^{-1}|k_2|\gg\cdots\notag\\
\text{or, }&\omega|k_{-1}|\gg\omega^0|k_0|\gg\omega^{-1}|k_1|\gg\omega^{-2}|k_2|\gg\cdots
\end{align}
 For order $n$, we have:
 \begin{equation}
\label{eqn70}
\omega^{-n}|k_n(r)|\gg\omega^{-(n+1)}|k_{(n+1)}(r)|.
 \end{equation}
 Since, the series converges, therefore the travelling waves have finite amplitude and are stable under linear perturbations.

\section{Asymptotic Sound Speed from Wave Equation}
\label{secVII}
In the asymptotic limit, we have:
\begin{equation}
\label{eqn71}
g^{tt}_{(0)}\big|_{r\to\infty} = \frac{1}{\rho_\infty r^2},
\end{equation}
\begin{equation}
\label{eqn72}
g^{tr}_{(0)}\big|_{r\to\infty} = g^{rt}_{(0)}\big|_{r\to\infty}=0
\end{equation}

\begin{equation}
\label{eqn73}
\text{and}\quad g^{rr}_{(0)}\big|_{r\to\infty} =-\frac{c_{s\infty}^2}{\rho_\infty r^2}.
\end{equation}
Hence, we get:
\begin{equation}
\label{eqn74}
c_{s\infty}^2=\left|\frac{g^{rr}_{(0)}}{g^{tt}_{(0)}}\right|_{r\to\infty}.
\end{equation}
The above result is similar to what one obtains for flat spacetime using the following wave equation for a scalar field, $\Psi(r,t)$:
\begin{equation}
\label{eqn75}
-\frac{1}{c^2}\partial^2_t\Psi+\partial^2_r\Psi=0.
\end{equation}

\section{Graphical Analysis}
\label{secVIII}
We consider the case of a spherically symmetric adiabatic flow satisfying the relativistic EoS given by Eq.~\eqref{eqn9}, accreting due to a Newtonian potential $\Phi = -\frac{1}{r}$. The sound speed is $c_s$ is defined by Eq.~\eqref{eqn13}. Bondi \cite{Bondi_1952} demonstrated that there exists a unique transonic solution of Eqs.~\eqref{eqn17} and \eqref{eqn18} which passes through a critical point $v_0(r)=c_{s0}(r)$. The solution can be derived by using the following equation for the Bernoulli constant:
\begin{equation}
\label{eqn75}
\mathcal{E} = \frac{v^2}{2}+h-\frac{1}{r}.
\end{equation}

\begin{figure*}[htb]
\centering
\includegraphics[scale=0.55]{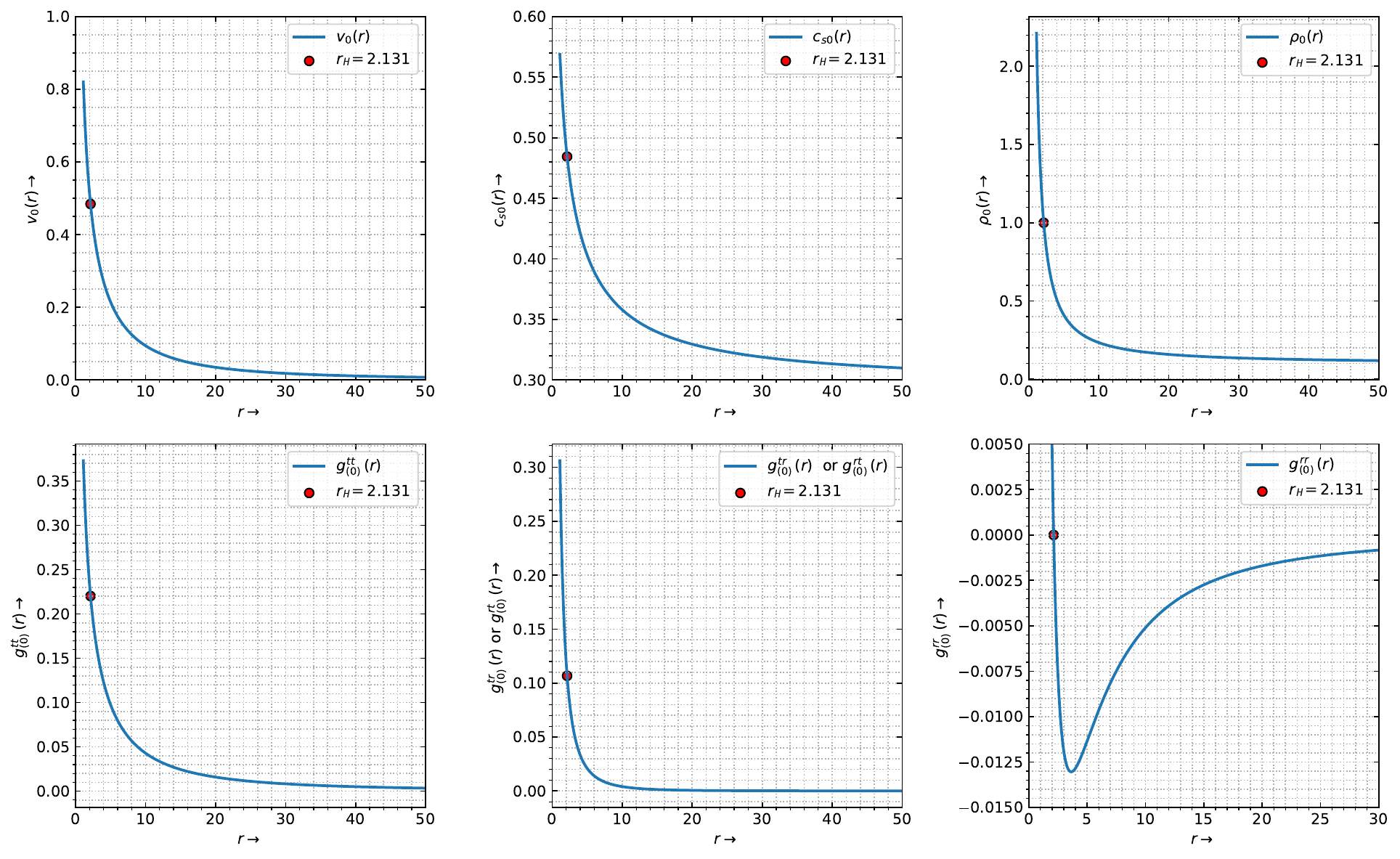}
\caption{The top panels are solutions for $v_0(r)$ (left), $c_{s0}(r)$ (middle) and $\rho_0(r)$ (right), with $f_0 = 2.199$. These solutions define the inverse acoustic metric components $g^{\mu\nu}_{(0)}$ plotted in the bottom panels: $g^{tt}_{(0)}(r)$ (left), $g^{tr}_{(0)}(r)$ or $g^{rt}_{(0)}(r)$ (middle) and $g^{rr}_{(0)}(r)$ (right). We find that $g^{rr}_{(0)} = 0$ at $r = r_H = 2.131$.}
\label{fig1}
\end{figure*}

We assume some representative values: $\mathcal{E} = 1.2$, $\xi = 1$ and $\eta = 1/1836$. With the accretor taken to be a black hole, the fluid velocity can approach close to the speed of light ($v=1$) at the inner radius. The characteristic plots of the background solution are in the top panel of Fig.~\eqref{fig1}. The solution has a mass accretion rate $f_0 = 2.199$. Using the $\rho_0(r)$, $c_{s0}(r)$ and $f_0$ solutions, we find the lowest order inverse acoustic metric components plotted in the bottom panel of Fig.~\eqref{fig1}, with the acoustic horizon at $r_H = 2.131$. It can also be seen that $g^{rr}_{(0)}>0$ for $r<r_H$ and $g^{rr}_{(0)}<0$ for $r>r_0$. However, $g^{tt}_{(0)}$ and $g^{tr}_{(0)}$ remain positive definite throughout the flow. At the acoustic horizon the dimensionless temperature $\Theta_H=152.263$.\\

We shall now check the consistency of these results. Let us consider the case of Sagittarius A* (Sgr A*), the supermassive black hole located at the centre of the Milky Way. The current best estimate of its mass $M$ is $4.297\times 10^6 M_\odot$. Its Schwarzschild radius $r_\text{Sch.}=\dfrac{2GM} {c^2}\approx1.274\times10^{10}\,\unit{m}$. Using our results, its acoustic horizon for the stationary background is located at $r_H=\dfrac{2.131GM} {c^2}\approx1.357\times10^{10}\,\unit{m}$. Now, $c_s<c$ suggests that $r_H>r_\text{Sch.}$. Hence, our estimate of $r_H$ for Sgr A* is consistent.\\

As $r_H$ and $r_\text{Sch.}$ are located very close to each other, it is quite reasonable to expect that the plasma temperature at the acoustic horizon and the event horizon would be of comparable order. The absolute temperature at the acoustic horizon [using Eq.~\eqref{eqn7}] is $T_H=\dfrac{m_{e^-}c^2\Theta_H}{k_B}\approx 9.036\times10^{11}\,\unit{K}$. This temperature is in agreement with the ion temperature range ($10^{11}–10^{12}\,\unit{K}$) expected in radiatively inefficient accretion flows around Sgr A*. Observations, including those from the Event Horizon Telescope, primarily constrain the electron temperature ($\sim10^9–10^{10}\,\unit{K}$), while ions are predicted to be significantly hotter. Accordingly, the temperature in our model should be interpreted as an effective (ion-dominated) fluid temperature, and the agreement should be understood at the order-of-magnitude level.

\section{Conclusion}
\label{secIX}
Only the expressions related to $g^{rr}$ change if one non-linearly perturbs a system with relativistic EOS instead of a non-relativistic one. This occurs due to the presence of $c_s$ in $g_{rr}$. $c_s$ and $\Gamma$ depend explicitly on $\Theta$ for relativistic EOS. Horizon variations depend on both $\Theta$ and $\rho$. $\Theta$ can have any arbitrary temporal dependency in general. The stability of stationary solutions for a general order of perturbation is inconclusive. This is due to the presence of non-zero time-dependent terms in general. Linear perturbations with relativistic EOS are stable for both standing and travelling waves. In particular, the travelling waves have two types of components --- phase related and amplitude related. The results obtained from the plots related to the stationary background metric are in agreement with the observational data.
\begin{acknowledgments}
SG acknowleges the RG HRI for extending support for academic vist after completion of the post-doc tenure. RGand BR acknowledges HRI for supporting a visit through the Research Apex Subproject Astro, during which he learnt the basics of the problem.
\end{acknowledgments}
\providecommand{\noopsort}[1]{}\providecommand{\singleletter}[1]{#1}%
%

\end{document}